# Some Historical Misconceptions and Inaccuracies Regarding The 1908 Tunguska Event

Andrei Ol'khovatov

https://orcid.org/0000-0002-6043-9205

Independent researcher
(Retired physicist)

Russia, Moscow
email: olkhov@mail.ru

**Dedicated to the blessed memory of my grandmother ( Tuzlukova Anna Ivanovna ) and my mother ( Ol'khovatova Olga Leonidovna )**

**Abstract.** This paper is a continuation of a series of works, devoted to various aspects of the 1908 Tunguska event. A large number of hypotheses about its causes have been put forward already. However, so far none of them has received convincing evidence. This is probably why new hypotheses appear almost every year, not only in the mass-media, but also in scientific literature. At the same time, any hypothesis should not contradict the known facts about the event. Unfortunately, the authors of new hypotheses, as well as the authors of popular science articles, often use data, many of which turned out to be not entirely accurate, or even incorrect. In this paper some of this data will be considered. Also the history of the Tunguska research is considered in this paper. Some other aspects of the 1908 Tunguska event are considered too.

## 1. Introduction

This paper is a continuation of a series of works in English, devoted to various aspects of the 1908 Tunguska event [Ol'khovatov, 2003; 2020a; 2020b; 2021; 2022; 2023a; 2023b; 2025a; 2025b].

More than a century has passed since the Tunguska event. A large number of hypotheses about its causes have been put forward already. However, so far none of them has received convincing evidence. As it is written on the title web-page of the

web-site created by KSE (see below about KSE) tunguska.tsc.ru/ru/ (translated by A.O.):

> "About a thousand researchers have devoted years of their lives to the Tunguska phenomenon. However, there is still no well-founded scientific understanding of what happened over the Siberian taiga on June 30, 1908."

This is probably why new hypotheses appear almost every year, not only in the mass-media, but also in scientific literature. At the same time, any hypothesis should not contradict the known facts about the event. Unfortunately, the authors of new hypotheses, as well as the authors of popular science articles, often use data, many of which turned out to be not entirely accurate, or even incorrect. In [Boslough and Bruno, 2025] an attempt was made to debunk some misunderstandings associated with the Tunguska event. However, when debunking some misunderstandings, it is advisable to avoid creating new misunderstandings, which is usually associated with mixing facts and interpretations of the Tunguska event (i.e. unconfirmed hypotheses still...). Therefore, this paper will focus only on the facts.

The Committee on Meteorites of the USSR Academy of Sciences (KMET) stopped research the area of the Tunguska event in the early 1960s. Later amateurs most of whom united under the name Kompleksnaya Samodeyatel'naya Ekspeditsiya (KSE) continued research. Since the late 1980s foreign scientists take part too.

Please pay attention that so called the epicenter of the Tunguska forestfall (the forestfall is named "Kulikovskii") is assigned to 60°53' N, 101°54' E. While discussing 1908-records from Russia, it should be remembered that according to the old Russian calendar the Tunguska event took place on June 17.

The author of this paper (i.e. A.O.) for brevity is named as "the Author".

## 2. Some historical misconceptions and inaccuracies

Let's start with a mistake in the Krinov's book [Krinov, 1966]. On page 127 E.L. Krinov presented translation of a text from a newspaper "Sibir" of 2 July 1908

> "At the beginning of the ninth hour in the morning of 17 June, a most unusual phenomenon of nature was observed here. In the village of Nizhne-Karelinsk (200 versts to the north of Kirensk) in the north west quite high above the horizon, the peasants saw a body shining very brightly (too bright for the naked eye) with a bluish-white light. It moved vertically downwards for about ten minutes. The body was in the form of "a pipe", i.e. cylindrical. …"

But the original newspaper article stated that Nizhne-Karelinsk is 30 versts to the north of Kirensk (which is much closer to reality) – see Fig.1.

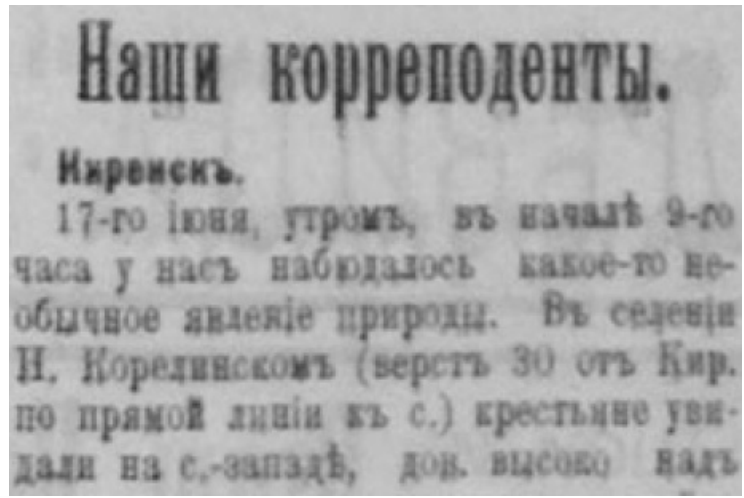

**Fig.1**

It is not clear how "30" transformed into "200" as in his book on Tunguska published in Russian in 1949 Krinov also wrote "200".

There is an interesting fragment from an account ([Vasil'ev, et al., 1981], translated by A.O.):

"A member of the management of the Consumers Association of Kezhma, I. K. Vologzhin, reported the following to L.A. Kulik on the 21st of November, 1921: "In the year 1908, on June 17, I was about 20 versts above the village Kezhma (along the Angara river) in the place Chirida. In the morning we examined our {probably fishing - A.O.} nets. It was a clear morning. There was not a cloud. With me was an old man. We heard - a thunder, then - another one, and gradually the thunder rumblings began to diminish toward the north. I don't remember what time it was, but the sun hadn't risen yet, it was just dawn. In the winter the inhabitants of the village Kezhma, who traded around the Podkamennaya Tunguska...""

Here is a fragment of a scan from [Vasil'ev, et al., 1981] with this account on Fig.2.

**Fig.2**

It follows from the account that something peculiar took place in the region several hours before the Tunguska explosion. However unfortunately L. Kulik in his article [Kulik, 1927] omitted the phrase "I don't remember what time it was, but the sun hadn't risen yet, it was just dawn.", and did not mark this with a relevant sign - see a scanned fragment from [Kulik, 1927] on Fig.3.

**Fig.3**

Sometimes the problem is an ambiguous description. For example, here is from [Longo, 2007]:

"In the early morning of 30 th  June 1908, a powerful explosion over the basin of the Podkamennaya Tunguska River (Central Siberia), devastated 2 150 ± 50 km$^2$  of Siberian taiga. Eighty millions trees were flattened, a great number of trees and bushes were burnt in a large part of the explosion area. Eyewitnesses described the flight of a "fire ball, bright as the sun"."

The words about "Eighty millions trees were flattened" were already considered in [Boslough and Bruno, 2025]. How to understand the words "devastated 2 150 ± 50 km$^2$"? Do the words mean complete destruction of the forest in this area? No, they don't. In reality the words mean that the area of the region, where at least a small fall of trees (assigned to the Tunguska event) has been noticed, is 2 150 ± 50 km$^2$. Here is what N.V. Vasil'ev (his surname sometimes is translated as Vasilyev) wrote [Vasilyev, 1998]:

"Around the epicentre there was a so-called "telegraph forest" area 3-5 km in dia, where almost all dead trees remained standing upright like "telegraph poles" with their crowns torn off. Outside of the limits of this area trees had fallen completely, then partially (Fast et al., 1967a, b, 1983)."

There is a good illustration of this fact on Fig.2 in [Robertson and Mathias, 2019]  (based on W. Fast data).

In the mid-1960s J. F. Anfinogenov made a map of completely devastated area (the map is based on aerial photo-survey conducted no later than 1949).  Basing on the map V.K. Zhuravlev calculated that the corresponding area is 500 square kilometers [Zhuravlev, 2012]. This fact should be taken into account when calculating the number of fallen trees.

Remarkably many trees survived near the epicenter – here is from [Vasil'ev, 2004] translated by A.O.:

"This impression is enhanced when looking at the map of the location in the area of the epicenter of the Tunguska explosion of trees that survived the disaster of 1908 (fig. 33, 34). It follows from it that in the Meteorite Basin, within a radius of five to six kilometers around the epicenter, there are at least sixty groups, and even as many as groves of trees that survived the 1908 disaster. Usually the trees are larch, much less often pine and spruce, however, also cedar (which is very sensitive to damage), and almost in the immediate vicinity of the calculated center of the disaster. These trees and groves are found on hillsides, in hollows/dells, and on the edge of swamps, and some <<living witnesses>> are found even in open spaces of swamps, i.e. in places that are not shielded at all."

It is important to note that there were even many survived groves of trees near the epicenter. Here is from [Boyarkina, et. al., 1964] (translated by A.O.):

"Of considerable interest is the presence and distribution of living old trees.  The epicentral zone contains a significant number of individual old trees and entire groves, which was discovered back in 1959-1960 [4, 12].
In the diagram [8, FIG.  6] all notable groves of old trees are marked. The largest of them are:  on the slope of the Wulfing (500 X 700 m), on the southern shore of the Southern Swamp (200 X 700 m), an array on the Western peat bog, a 200 X 300 m, groves to the east of the swamp, etc."

Here is what I.K. Doroshin wrote about the Wulfing grove [Doroshin, 2005] (translated by A. O.):

"...at least one grove consisting of larch, spruce and cedar (near the Wulfing mountain). In such places, the trees either did not lose their crowns at all, or the loss was minimal. There are either no traces of the 1908 fire here at all (a grove near the Wulfing mountain), or there are traces of a grass-roots fire of varying intensity;... ".

The Wulfing grove is situated about 3 km to the north-west from the epicenter.
Therefore, the comparison of the destruction in Tunguska with the explosion of a powerful hydrogen bomb, which is widespread in the media, is not entirely correct.
More on the peculiarities of the epicentral area effects can be read in [Ol'khovatov, 2025b].
It is worth to add that residents reported about other forestfalls possibly occurred together with the Kulikovskii one - see [Ol'khovatov, 2021]. However areas of these forestfalls seem to be much less than the area of the Kulikovskii forestfall. Among them, only the Chuvar forestfall was researched in the 1960s-1970s.
By the way, the Chuvar forestfall is a remarkable one as trees fell in opposite direction of the expected direction. Here is a fragment from [Vasil'ev et al., 1960] (translated by A.O.) about KSE discovery of the Chuvar forestfall:

"And suddenly the terrain landscape changed dramatically. The eastern slope was quite steep, and here, on this slope, the group again entered the forestfall zone, and even what the forestfall! Such a picture none of the three did see either on the Makikta-river, or to the south of the Hushma river, or on the path of Kulik. The landscape seemed fantastic: huge trees, almost up to a meter across, were uprooted, thrown on top of each other, split like matches."

The fallen trees were with their tops towards the epicenter (see more on the Chuvar forestfall below).

It is noteworthy that according to the memoirs of Nikolai Fedorov (a young painter who took part in the Kulik last expedition in 1939), Kulik sent him and his partner (on their way to Vanavara) to inspect another forestfall in the area of the Tatare river [Kulik – Pavskii, 2003]. However, due to bad weather, it was not possible to get to the forestfall. Unfortunately, the Author knows nothing more about this forest-fall. Very approximately, we can talk about the forest-fall at a distance of ~100-150 km SE from the epicenter.

Now about the number of deaths caused by the Tunguska event. The Author agrees with [Jenniskens et al., 2019] that the number of more or less firmly established deaths is 3. Also in [Jenniskens et al., 2019] there is a section "3.4. Possible other casualty reports". The Author would like to add one more hint regarding more deaths from the Tunguska event.

The author of the lines below was Aleksei Smirnov - an experienced traveler, a hunter, a local historian and a writer. In 1928, he participated in a rescue expedition aimed at helping/saving L.A. Kulik. While hunting he lost his way in the region of the fallen forest. Here is from [Smirnov, 1929] (translated by A.O.):

"I sit down on the trunk of a tree and light a cigarette, but it immediately loses its taste. A white human skull is grinning at me from under my feet!.. Who found death here and when? But the dead taiga will never reveal this secret ...

An unexpected find is unpleasant. "What if I break my leg here? Who will find me then?" - the thought flashes. The environment is depressing. Without finishing my cigarettes, I leave the trunk and move on."

In [Jenniskens et al., 2019] there are accounts of Evenks about the Chuvar ridge (the Chuvar forestfall):

"... All the fallen trees fell with their tops to the East. The storm was so strong, that even the chum broke and was blown away."

The authors of [Jenniskens et al., 2019] commented the accounts:

"On Chuvar Ridge, the trees fell to the West, so perhaps the "tops" refers to the exposed roots, or the direction was mistakenly reversed. <...> The location marked "7" is at a place on the Chuvar ridge where trees would have fallen in an east to west direction."

However in reality Evenks were correct - the trees fell with their tops to the East on the Chuvar ridge! This fact was established by KSE already in 1959.

On Fig.4 here is a drawing of the fallen trees directions in the trial areas from [Lyskovskii, 1999]. A grid of special coordinates [Fast, et al., 1976] is plotted, a bold dot indicates the calculated epicenter of the Tunguska explosion. The points are trial areas, and the vectors from these points are the average directions of the fallen trees on these trial areas. Coordinates (x,y) are given in km. Axis "x" is directed to the magnetic north (see details in [Fast, et al., 1976]). The trial areas near x=40 km, y~0 km belong to the Chuvar forestfall.

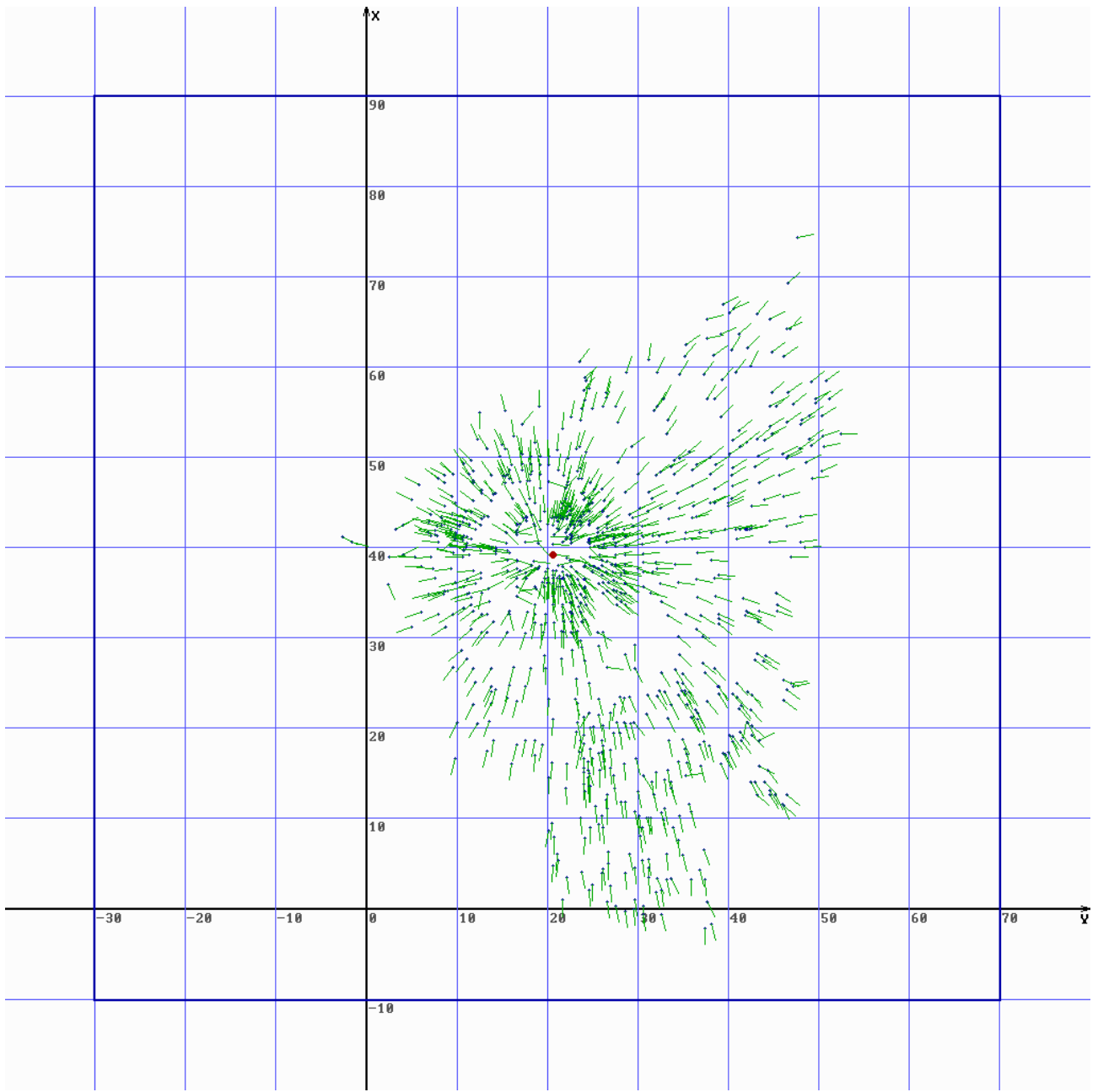

**Fig.4**

In 1971 the Chuvar forestfall was mapped. The map is shown on Fig.5 (see http://tunguska.tsc.ru/ru/archive/cae/1947/48-80/57/ for details). On Fig.5 marks 1,2 are areas of the most strong forestfall, mark 3 - less intensive. According to the 1971 research this forestfall has a length of 7000 m, and a width of 1500 m. The boundaries of the forestfall are rather sharp.

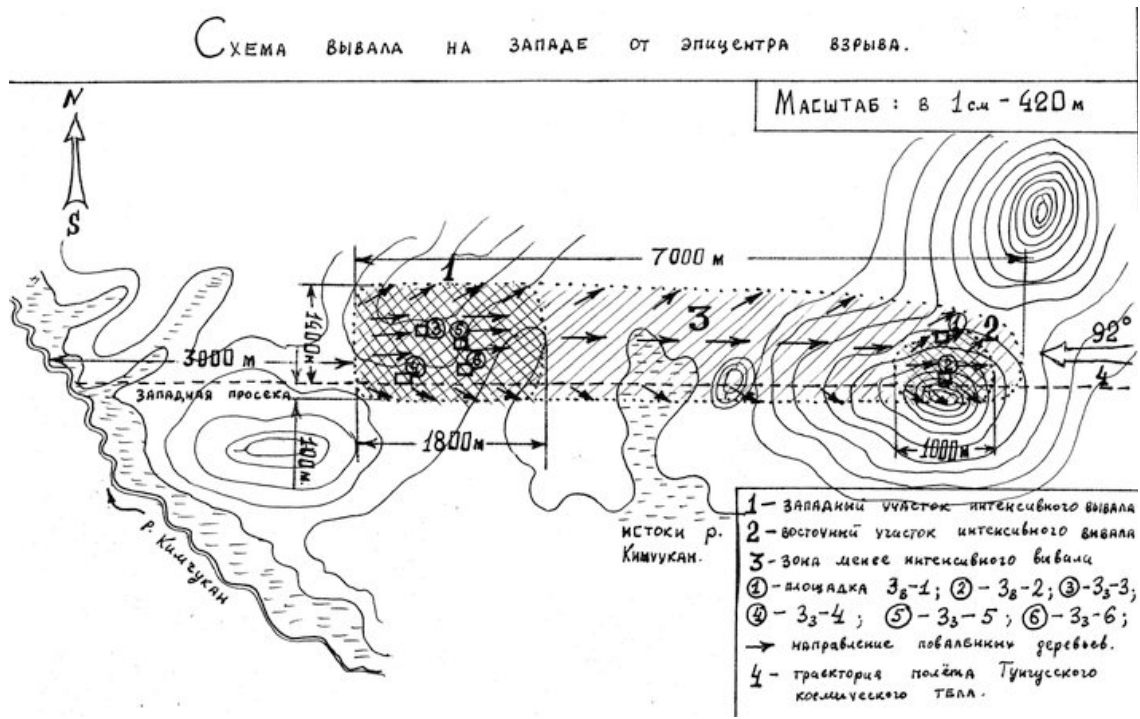

**Fig.5**

The Author summarized the available data on the Chuvar forest-fall in [Ol'khovatov, 2021].

Let's continue about misconceptions. It was written on page 13 of the book [Bruno, 2022]:

"A red glow surrounded Ivan Aksenov on a summer morning in 1908 as he carved the carcass of an elk he had just hunted with his now crimson-colored knife. Looking up in fear, he saw a fiery devil flying toward him. As he started to pray, a powerful gust knocked him unconscious. Aksenov woke to find a blazing forest falling around him. Upon returning to camp, he decided to go out with some others to try to extinguish the fire."

Here is from the Alsenov's original account from [Vasil'ev et al., 1981] (translated by A.O.):

"<…> Aksenov was 24 years old. Early in the morning, Aksenov went hunting for a moose, shot a moose on the Chamba {river –A.O.} somewhere above the mouth of the Makikta {river-A.O.} and began to skin the carcass. When he was working (bending over the carcass) "suddenly everything turned red." He was scared, threw up his head - "and at that moment it hit," and he lost consciousness for a while. "When I woke up, I saw it falling all around, burning. Don't believe, Viktor Grigoryevich [V.G. Konenkin], that God was flying there, the devil was flying there. I raised my head and saw the devil flying. The devil himself was like a chock, light in color, two eyes in front, fire behind. I got scared, I closed up with his clothes on, and began to pray. (I did not pray to a pagan god, I prayed to Jesus Christ and the Virgin Mary). I prayed and woke up, nothing happened anymore. <…>".

So Aksenov said that he saw the "devil" after the explosion! Here is a fragment of his account in Russian from [Vasil'ev et al., 1981] on Fig.6.

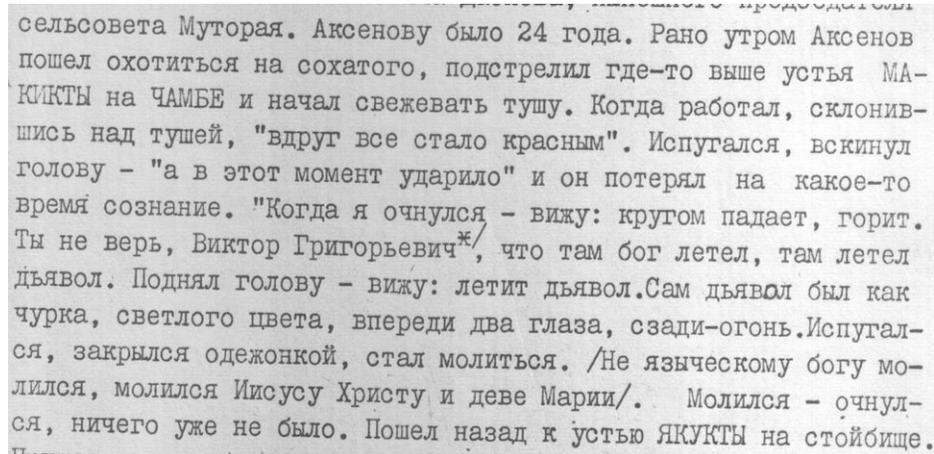

**Fig.6**

Vasil'ev wrote [Vasilyev, 1994]:

"A particular feature of Aksenov's account (agreeing otherwise with the early evidence of Vanavara residents that Kulik had heard as far back as the 20s), is the assertion of the eye-witness that after the explosion he had seen an object flying down the Chamba, i.e. generally north to south. He called the object a "devil"."

The Aksenov's account surprised many of the Tunguska researchers. Indeed, what was the strange object that flew some time after the explosion?.. More info on the Aksenov's account can be found in [Ol'khovatov, 2023b].

Ilya Potapovich Lyuchetkan reported in Vanavara in 1935 [Vasil'ev et al., 1981], translated by A.O.:

"[I] saw. [It] flew low-low across the sky over the forest, and shot very, very often. And when it fell, it shot even louder..."

Please note that Lyuchetkan is talking about another low-flying object, and, by the way, as in the case of the Aksenov's object, of apparently moderate brightness. More on similar "objects" can be read in [Ol'khovatov, 2023a; 2023b].

In view of the above, let's consider a statement from [Longo, 2007] that eyewitnesses described the flight of a "fire ball, bright as the sun". Here is what some other researchers wrote. Here is, for example from [Zotkin and Chigorin, 1988] translated by A.O.:

"As you know, eyewitnesses claim that it was in the order of magnitude slightly weaker than the Sun.  The definitions of the intensity of the luminosity in this case have a very large uncertainty, however apparently, the estimation the stellar magnitude -22$^m$ ÷ -17$^m$  will be reasonable."

Similar opinion had a prominent Tunguska researcher V.A. Bronshten – here is from [Bronshten, 2000], translated by A.O.

"The effective temperature of the Tunguska plasmoid, judging by the accounts of eyewitnesses, was slightly lower than the solar temperature (it was possible to look at the Tunguska bolide, it did not dazzle eyes)*."

Indeed, if to look at the most reliable accounts (which were collected in 1908), then it is possible to see that almost all of them did not mark the luminous phenomena as being very bright (the accounts were translated in [Ol'khovatov, 2023a]). Among them only one report said about very high brightness of the body. But the body was a remarkable one. It was the body described in the text from the newpaper "Sibir" of 2 July 1908 which was presented in [Krinov, 1966], and it is mentioned at the beginning of this paper. This long-living (10 minutes!) fiery pillar was not a fire-ball anyway. Remarkably this body was reported only from two closely located settlements.

By the way, it looks like some pillars of fire originated from a fireball were reported – here is from an account collected in 1908 [Ol'khovatov, 2023a]:

"...a fiery ball appeared, which flew in the direction from the southeast to the northwest. This ball, approaching the earth, took the form of a flattened ball from above and below (as it was visible to the eye); approaching even closer to the earth, this ball had the appearance of two pillars of fire."

Indeed, eyewitness accounts indicate that the Tunguska event was associated with many pillars of fire – see [Ol'khovatov, 2023b], as well as "low-flying objects"...

There is another remarkable aspect. Here is from [Demin et al., 1984] where an analysis of all known accounts was conducted (translated by A.O.):

"The relative frequency of visual observations as a function of distance from the epicenter is also quite interesting. Most observations are grouped at distances of 400-500 and 800-1000 km, and in a circle with a radius of 200 km, with 35 accounts related to other aspects of the phenomenon, there are practically no visual observations of the flight."

Indeed there is a couple of accounts (by Aksenov, and Lyuchetkan) about low-flying objects with moderate brightness. There is the only report of a very bright flying object. Evenk Nikolai A. Kocheni (born in 1880) observed (and reported in 1934) "a flying round-headed fire arrow, and it has a tail of feathers sticking out from behind" from the South Chunya river (~100 km to NNE from the epicenter) and said that his eyes hurt looking at it. However, none of the eyewitnesses (interviewed in the 1960s) in this region confirmed such an observation (while they reported about conditions of the sky, etc.). Moreover, in 1932 a geophysicist S. Ovchinnikov visited the epicenter, and collected some Evenki accounts in the region. In 1934 he wrote to L.A. Kulik [Vasil'ev et al., 1981] (translated by A.O.):

"The main area of my stay was the areas located significantly north of its location {i.e. the epicenter - A.O.}, and those groups of Evenks have no information about its appearance and fall.
    Of course, you know the very general stories that the Chunya's Evenks told me."

So Kocheni seems to be the only eyewitness who saw this bright object in the area. It hints that the object was of local origin and at low-altitude.

It can't be ruled out that possibly similar bright object was seen in Sulomay (about 577 km to the west from the epicenter). The flight in Sulomay was accompanied with a strong wind which put down some tall trees [Ol'khovatov, 2020a].

In any case, it is correct to say that the vast majority of eyewitnesses reported that the brightness of the object was moderate.

It is possible to add that it is noteworthy that none of the most reliable eyewitnesses saw even just a trail of the alleged super-bolide.

Now let's consider so called bright nights which took place near June 30, 1908.

In general there were various optical sky peculiarities at those times.

Krinov wrote [Krinov, 1966]:

"On the first night after the fall of the Tunguska meteorite, i.e. from 30 June to 1 July 1908, and with lesser intensity on a few successive nights, extraordinary optical phenomena were observed in the Earth's atmosphere."

The late evening of June 30, 1908 was remembered by eyewitnesses, first of all, because of the colorful twilight, which in some places lasted all night. So D.D. Rudnev in the village of Muratovo (the Oryol province) captured the twilight with his photocamera at 0:30 on July 1, and then manually colored the picture. The shutter speed was 10 minutes. This picture is shown in Fig.7 taken from [Nadein, 1909]. Since Muratovo is a very common name for a settlement, the Author can only say that it is most likely the village of Muratovo with coordinates about 53.2° N, 35.8° E.

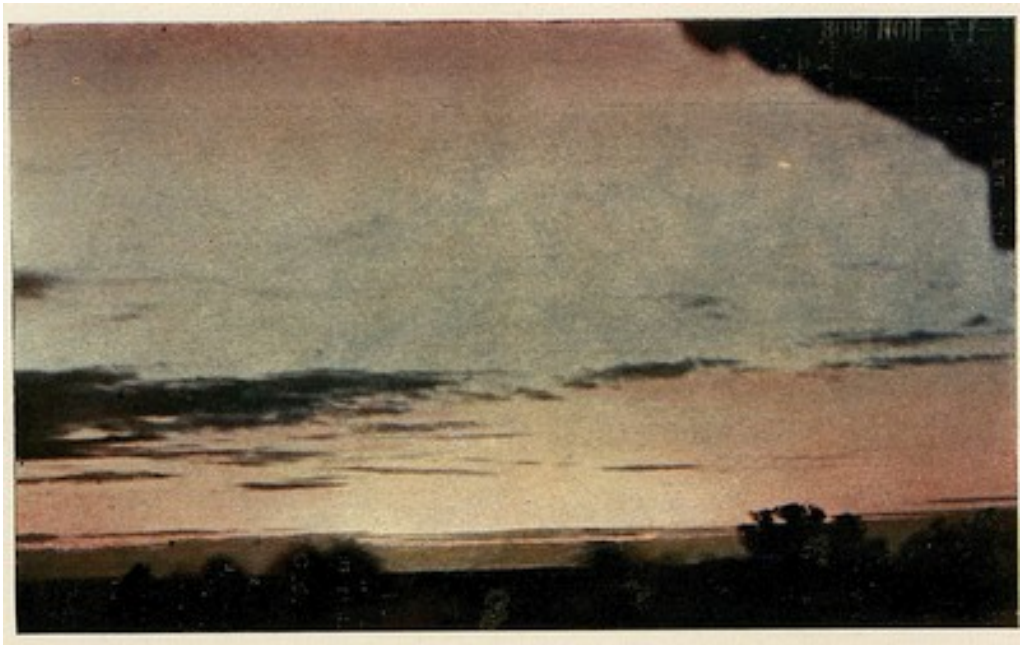

**Fig.7**

However the phenomena commenced before June 30. Here is from [Vasilyev, 1998]:

"A detailed analysis of dynamics of June-July 1908 optical anomalies gives ground to a supposition that their first signs had appeared already several days before the fall of the meteorite: Suring (1908) points out that they began on 23 June, F. de Roy (1908) about 25 June and Denning (1908a, b) on 29 June. On that last day they were registered at 8 points in Germany, Holland, Britain, Sweden, Poland and Russia (Fig. 6). Nevertheless, the events reached their maximum (observed at more than 140 points) on the night of 30 June to 1 July.<…> Beginning with 1 July they vanished exponentially, but post-effects continued until the end of July 1908."

Here is what Alan Harris wrote [Harris, 1996]:

"Curiously, the anomalous "white nights" were reported as starting a full week before the impact, an observation that is hard to reconcile with even a comet since the Earth could not possibly spend a full week cruising through the tail. More likely, unusually intense auroral or noctilucent cloud activity unrelated to the impact occurred before the event."

In the Author's opinion the sky optical anomalies and the explosive event in Tunguska are related. The Author wrote a paper on bright nights and the Tunguska event [Ol'khovatov, 2025a].

## 3. Conclusion

The presented material is only the tip of the iceberg, but it points that the Tunguska event was a very complex phenomenon. A reader is welcome to make his own conclusions regarding the Tunguska event. Anyway, any interpretation of the Tunguska event should explain its various manifestations as well as accounts.

**ACKNOWLEDGEMENTS**

The author wants to thank the many people who helped him to work on this paper, and special gratitude to his mother  - Ol'khovatova Olga Leonidovna (unfortunately she didn't live long enough to see this paper published...), without her moral and other diverse support this paper would hardly have been written.